\documentclass[12pt,a4paper]{article}
\usepackage[english]{babel}
\usepackage[utf8x]{inputenc}
\usepackage{ucs}
\usepackage{amsmath}
\usepackage{amssymb}
\usepackage{graphicx}
\usepackage[super,comma,sort&compress]{natbib}
\usepackage{bm}
\usepackage{color}
\usepackage{soulutf8}
\usepackage{Sweave}
\usepackage{url}

\usepackage{authblk}

\begin{document}

\title{Obtaining adjusted prevalence ratios from logistic regression model in cross-sectional studies}
\author[1]{Leonardo Soares Bastos}
\author[2]{Raquel de Vasconcellos Carvalhaes de Oliveira}
\author[3]{Luciane de Souza Velasque}
\affil[1]{Scientific Computing Program, Oswaldo Cruz Foundation, Brazil}
\affil[2]{ Evandro Chagas National Institute of Infectious Diseases, Oswaldo Cruz Foundation, Brazil}
\affil[3]{Department of Mathematics and Statistics, Federal University of State of Rio de Janeiro, Brazil}

\date{}
\maketitle

\begin{abstract}
In the last decades, it has been discussed the use of epidemiological prevalence ratio (PR) rather than odds ratio as a measure of association to be estimated in cross-sectional studies. The main difficulties in use of statistical models for the calculation of PR are convergence problems, availability of adequate tools and strong assumptions. The goal of this study is to illustrate how to estimate PR and its confidence interval directly from logistic regression estimates. We present three examples and compare the adjusted estimates of PR with the estimates obtained by use of log-binomial, robust Poisson regression and adjusted prevalence odds ratio (POR). The marginal and conditional prevalence ratios estimated from logistic regression showed the following advantages: no numerical instability; simple to implement in a statistical software; and assumes the adequate probability distribution for the outcome.
\end{abstract}


\section*{Introduction}

During the last decades, several authors\cite{axelson1994,stromberg1994, stromberg1995, skov1998, barros2003, deddens2003, coutinho2008} have been studying the best association measure to be estimated in cross-sectional studies. The consensus is that the prevalence odds ratio (POR) is a good approximation of the prevalence ratio (PR) \cite{kkm1982epidemiologic}, if and only if, it is in the presence of a rare event. Logistic regression is the most popular statistical model used when estimating POR due to ease of interpretation and computational implementation. However, when the choice of association measure is the PR, this model produces poor estimates in the presence of a not rare event. In such context, several authors proposed alternatives methods instead of using logistic regression to estimate the true PR.

Lee~\cite{lee1993} is one of the first authors to use Cox models with Breslow's modification (Breslow-Cox model) to estimate prevalence ratios, yet their standard errors and, consequently, confidence intervals are not correct. Although the correction for standard errors obtained by Cox models has already been proposed~\cite{lin1989robust}, Lee~\cite{lee1993} did not correct them. Barros and Hirakata\cite{barros2003} use the fact that Breslow-Cox and Poisson models estimate the same effects\cite{clayton1993statistical} and use Poisson regression models with robust variance to estimate the PR. Zou (2004)~\cite{Zou2004} provided a simulation study demonstrating the reliability of the Poisson model with robust variance to estimate PR in tables 2 by 2. The main problem to use Poisson model while estimating PR is the misuse of a specific counting probability distribution to describe a response variable that is dichotomous (presence or absence of an outcome).

Skov et al. (1998)\cite{skov1998} used a generalized linear model with the binomial distribution and log link (log-binomial model) to estimate directly PR~\cite{wacholder1986binomial}. Although this model makes possible to estimate directly the PR and assumes the appropriated probability distribution considering the type of the variable response, the lack of convergence in the presence of continuous variables is still a problem. For solving this problem, Deddens~\cite{deddens2003} introduced the COPY method to find an approximation to the MLE when the log-binomial model fails to converge. Due to the convergence problem of the log-binomial model, Schouten et al.~\cite{schouten1993} proposed a simple data manipulation in order to use the logistic regression to obtain the PR. It consists in modifying the data set by duplicating the lines where the event occur and replacing the outcome from event to non-event~\cite{schouten1993,lumley2006,diaz2012simple}.

Other approach was proposed by Wilcosky and Chambless (1985)~\cite{wilcosky1985comparison}, using the conditional and marginal methods~\cite{lin1989robust}, which developed a direct adjustment of epidemiological measures from binary regression. An advantage is that this method assumes a probability distribution for variable with binomial response, which matches the nature of the observed variable as response in cross-sectional studies. We find one article18 that uses the Wilcosky and Chambless~\cite{wilcosky1985comparison} method to estimate PR, yet it did not mention anything about the software implementation. Recently, Ospina and Amorim (2013)~\cite{OspinaAmorin2013} developed a package to R software (prLogistic) to estimate marginal and conditional PRs with bootstrap and delta's method confidence interval, but they were not shown the details of this package in scientific article as well as the differences between the several methods to estimate PR.

In this article we implement the direct approach to estimate the prevalence ratio from binary regression models based on Wilcosky and Chambless (1985)~\cite{wilcosky1985comparison} and compare with different methods to estimate the prevalence ratio presented in the literature. Three different data sets are used to illustrate our study.

\section*{Methods}

We use real and simulated data to compare prevalence ratio (PR) estimates obtained by the marginal and conditional models based on the approach proposed by Wilcosky and Chambless (1985)~\cite{wilcosky1985comparison}. Those estimates are also compared with the estimates obtained by the Binomial, Log-binomial and robust Poisson/Cox models.

It is well known that it is possible to estimate the probability of occurrence of a disease (denominated prevalence in transversal studies) adjusted for two or more variables across the logistic model. Suppose, for example, that one has information about diabetes status (1: Yes / 0: no), age (continuous) and obesity (1: Yes / 0: No) of a defined population, we can obtain the probability of diabetes by the equation below.
\begin{equation} \label{prob}
P(DIABETES = 1) = \frac{1}{1 + \exp\{ - (\beta_0 + \beta_1 AGE + \beta_2 OBESITY )\}}
\end{equation}
where are $\beta_0$, $\beta_1$ and $\beta_2$ are regression coefficients estimated from the data. Note that $\exp(\beta_2)$ estimates the odds ratio for diabetes in obese compared to non-obese, adjusted by age. However, if we are interested in obtaining the estimated PR for diabetes in obese and non-obese adjusted by age, we can proceed in two ways as described below: 

\begin{enumerate}
\item Marginal Model 

In each stratum of variable OBESITY (Yes or No), the diabetes prevalence is calculated for each age value observed in the dataset using equation (\ref{prob}). The PR is the ratio between the average of prevalences in each stratum. This estimate is called by Wilcosky and Chambless (1985)~\cite{wilcosky1985comparison} marginal prevalence ratio (MPR). 

\item Conditional Model

In each stratum of variable OBESITY, the diabetes prevalence is calculated using eq. 1 setting age as an average value obtaining from the dataset. Thus, the ratio of the two prevalences can be calculated. Wilcosky and Chambless (1985)~\cite{wilcosky1985comparison} named this method as conditional prevalence ratio (CPR).
\end{enumerate}

In the linear regression, both approaches estimate the same value. However, in the logistic model we observed significant differences between the estimates of the two models when p is close to zero or one. 

According to Lee~\cite{lee1993}, the marginal model provides an internally adjusted measure and this would undermine the comparability of external estimate of PR. While the conditional model you can use default values ​​as the average value of covariates and this allows comparisons with other populations of studies that used the same default value. More details about the marginal and conditional models can be found in Lee~\cite{lee1981covariance} and Wilcosky and Chambless (1985)~\cite{wilcosky1985comparison}.

Asymptotic confidence intervals for the conditional and marginal prevalence ratios were proposed by Flanders and Rhodes (1987)~\cite{flanders1987large}. The authors also presented a SAS code to estimate and calculate the intervals of the conditional and marginal prevalence ratio. In this paper, we implemented their functions 
and compared with other methods by applying in different data sets in the results section

The prevalence ratio estimation methods are illustrated in three different databases, all containing a binary outcome Y, a binary exposure X, and at least one controlling variable Z.

Application 1: The first database is a toy example with a 1000 simulated observations and one continuous controlling variable. In this example, we simulated 1000 binary outcomes with a binary exposure, X, and a continuous confounding variable, Z. The exposure was sampled from a Bernoulli distribution with probability 0.5, the confounding variable was sampled from a Normal distribution with mean zero and unit variance. The outcome was sampled from a Bernoulli distribution with probabilities such as the baseline prevalence equals 20\%, the conditional prevalence ratio for X at Z = 0 is equal 2, i.e. PRX|Z=0 = 0, and the conditional prevalence ratio for X at Z = 1 is such as that regression coefficient is $\beta_2 = 0.20$, hence PRX|Z=1 = 1.9186. There are several possible values for the conditional prevalence ratio for X depending on Z.

Application 2: The second database refers to a cohort of 1273 live births in 1993 in the city of Pelotas, Brazil, with the aim of linking social-demographic factors, and reproductive health, informed by the responsible female, to the nutritional condition of their children after 4-5 years\cite{barros2003}. The analysis included the underweight at 4-5 years as outcome, Y, whose prevalence was 4.1\%, previous to hospitalization as exposure, X, and birth weight (normal or low birth weight) as controlling variable, Z. For this application, we can calculate the prevalence ratio applying Mantel- Haenszel method, because all variables are binary. The Mantel-Haenszel method is considered here the gold standard method.

Application 3: The third database analyzes 703 women sexually active HIV-infected treated between 1996 and 2007 in Rio de Janeiro, Brazil, with no history of hysterectomy, in order to assess factors associated with high-grade squamous intraepithelial lesions (HSIL), lesions that can progress cancer of the cervix\cite{luz2012}. Five variables were analyzed: the presence of HPV as exposure variable, X, the cervical cytological abnormalities as outcome, Y, and three, Z, controlling variables (age, number of pregnancies and time since last gynecological examination in HIV-infected women). Variables X and Y are binary variables and the others are continuous variables. The prevalence of outcome was 4.1\%.

Adjusted prevalence ratios and prevalence odds ratio were estimated by several different methods. Prevalence ratio were estimated by robust Poisson and log-binomial models, the modified database Schouten et al. approach, and the conditional and marginal prevalence ratio proposed by Wilcosky and Chambless(1985)~\cite{wilcosky1985comparison}. Prevalence odds ratio (POR) using the usual logistic regression were also presented.

The different methods to obtain prevalence ratios were coded in R~\cite{R2014}. The code is available in the appendix. 

\section*{Results}

\subsection*{Application 1: Toy example}

Table \ref{ap1.tbl} presents different estimates for the prevalence ratio for variable X using different approaches. The crude prevalence ratio underestimates the prevalence ratio that varies from 1.71 to 2.25, whereas the crude and the adjusted prevalence odds ratio overestimate the prevalence ratio. The adjusted prevalence ratios are all very similar, providing reasonable estimates. The estimates differ only in the second or third decimal places, with the smallest estimated value in the log-binomial model and the largest in the conditional prevalence ratio.

\begin{table}[ht]
\caption{Prevalence ratios and respective 95\% confidence interval estimates in the analysis of the toy data using Y as the outcome, X as the risk factor and the continuous covariate Z as control factor.} \label{ap1.tbl}
\begin{center}
\begin{tabular}{lcc}
  \hline
            & PR & 95\% CI \\ \hline
Robust Poisson & 1.950 & (1.573, 2.416) \\ 
Log-binomial & 1.942 & (1.575, 2.418) \\ 
\multicolumn{2}{l}{Logistic regression} & \\ 
\multicolumn{1}{r}{POR} & 2.537 & (1.905, 3.398) \\ 
\multicolumn{1}{r}{$PR^{(c)}_{X}$} & 1.956 & (1.578, 2.425) \\ 
\multicolumn{1}{r}{$PR^{(m)}_{X}$}     & 1.949 & (1.574, 2.414) \\ \hline 
\end{tabular}
\end{center}
Note: The conditional prevalence ratio for X varies from 1.71 to 2.25 depending on the value of Z, POR = Prevalence odds ratio and PR = Prevalence ratio.
\end{table}

\subsection*{Application 2: Underweight in the 4-5 year old children in Pelotas-RS/Brazil}

Table \ref{ap2.tbl} presents the adjusted prevalence ratio of the occurrence of underweight in the 4-5 year old children (outcome) by previous hospitalization (exposure) controlled by birth weight (normal or low birth weight).

\begin{table}[ht]
\caption{Prevalence ratios (PR) and respective 95\% confidence interval estimates in the analysis of the data using underweight as the outcome (Y), previous hospitalization (X) as the risk factor and birth weight as control factor (Z).} \label{ap2.tbl}
\begin{center}
\begin{tabular}{lcc}
  \hline
            & PR & 95\% CI \\ \hline
MH PR       &  2.483 & (1.456, 4.235) \\            
Robust Poisson & 2.479 & (1.454, 4.226) \\ 
Log-binomial & 2.481 & (1.447, 4.226) \\ 
\multicolumn{2}{l}{Logistic regression} & \\
\multicolumn{1}{r}{POR} & 2.641 & (1.481, 4.671) \\ 
\multicolumn{1}{r}{$PR^{(c)}_{X}$ } & 2.532 & (1.471, 4.357) \\ 
\multicolumn{1}{r}{$PR^{(m)}_{X}$ } & 2.460 & (1.451, 4.171) \\ \hline 
\end{tabular}
\end{center}
Source: Data from Table 1 of Barros and Hirakata (2003)\cite{barros2003}\\
Note: POR=Prevalence odds ratio, PR=Prevalence ratio and MH PR =  Mantel-Haenszel adjusted Prevalence Ratio.
\end{table}

Despite the low prevalence of the outcome (4.1\%), a difference of 0.169 between the crude PR and the crude POR for the previous hospitalization is observed. According to the crude PR, for those children that were previously hospitalized have a larger prevalence (Crude PR = 2.902) of being underweight, when compared with those without previous hospitalization. The adjusted prevalence ratios of the log-binomial, robust Poisson, marginal prevalence ratio and Mantel-Haenszel approach presented similar estimates (2.481, 2.479, 2.460, and 2.483, respectively). The largest adjusted estimates were the POR (2.641) and the conditional prevalence ratio (2.532).

\subsection*{Application 3: Cervical cytological abnormalities in HIV-infected women}

Table \ref{ap3.tbl} shows the influence of high risk HPV (exposure) in the occurrence of cervical cytological abnormalities, controlled by age, number of pregnancies and time since last gynecological examination in HIV-infected women. Despite the low prevalence the crude POR differs from the crude PR by 0.6. Those women with high risk HPV present 640\% more cytological abnormalities. The adjusted POR is the highest estimated value (7.990). The adjusted prevalence ratios obtained by the log-binomial, robust Poisson approaches and the marginal prevalence ratio are very similar. The conditional prevalence ratio increases the ratio in up to 46\% when compared to the other adjusted methods.

\begin{table}[htp]
\caption{Prevalence ratios and respective 95\% confidence interval estimates in the analysis of cervical cytological abnormalities in HIV-infected women using high risk HPV as an exposure variable (X), and controlling by (Z) variables.} \label{ap3.tbl}
\begin{center}
\begin{tabular}{lcc}
  \hline
            & PR & 95\% CI \\ \hline
Robust Poisson &  7.123 & (2.489, 20.388) \\
Log-binomial & 7.192 & (2.849, 24.135) \\
\multicolumn{2}{l}{Logistic regression} & \\
\multicolumn{1}{r}{POR} & 7.990 & (3.029, 27.531) \\
\multicolumn{1}{r}{$PR^{(c)}_{X}$} & 7.529 & (2.617, 21.665) \\
\multicolumn{1}{r}{$PR^{(m)}_{X}$ }     & 7.118 & (2.518, 20.124) \\ \hline
\end{tabular}
\end{center}
Note: Z variables=age, number of pregnancies, and time since last gynecological examination; POR=Prevalence odds ratio and PR=Prevalence ratio.
\end{table}

\section*{Discussion}

Difficulties in obtaining prevalence ratio in cross-sectional studies have been investigated by several authors in recent years. Several authors use strategies for indirect calculation of the PR using the Breslow-Cox and Poisson models (with and without robust variance), while others interpret the prevalence odds ratio obtained in logistic regression models as prevalence ratio.
Lee~\cite{lee1993} is one of the first authors to discuss the methods proposed for estimating prevalence ratio. Most cross-sectional studies in health, until then used logistic regression model, since it has the advantage of adjusting for the effects (PORs) for several variables, either confounding or modifying effect, however, the POR can poorly estimate the prevalence ratio, up to 27 times more when the outcome is prevalent~\cite{lee1994odds}.
 
Regarding the estimation of adjusted prevalence ratio, in our examples the estimates provided by log-binomial model, robust Poisson model, Schouten et al. approach, and marginal prevalence ratio are similar. The conditional prevalence ratio (CPR) differs from the other estimates but it is still smaller than the adjusted POR. The CPR proposed by Wacholder~\cite{wacholder1986binomial} is the prevalence ratio conditional on the mean values of the covariates, yet one could condition on any value for the confounding variables (higher or lower risk scenario). For instance, the prevalence of cervical cytological abnormalities in HIV-infected women (Application 3) was estimated for each of the 703 women  based on their respective values for  age, number of pregnancies, and time since last gynecological examination (Z variables) in a set of presence high risk HPV (X=1) and the mean value of prevalence was computed. Similar calculations were performed in a set with absence of high risk HPV (X=0). For more detailed information conditional and marginal methods are well described in Wacholder~\cite{wacholder1986binomial} and Wilcosky and Chambless~\cite{wilcosky1985comparison}.

The main advantage of the log-binomial and robust Poisson models is that they are already implemented in the most popular statistical packages. The log-binomial has the disadvantage of not use a proper link function leading to numerical instability in the estimation process, resulting in non-convergence issues. The COPY method~\cite{deddens2003} was proposed to achieve convergence with the log-binomial model, but this method is only available in SAS, which is a proprietary software. The robust Poisson model assumes that all the events in the database occurred in the same time. Also, there is an inappropriate use of Poisson distribution by modeling a binary outcome~\cite{deddens2003}. However, it is important to highlight that the likelihood of the Poisson model has been used only to obtain an estimation equation and not for the purpose of modeling a binary response variable. The Schouten et al. approach~\cite{schouten1993} can be implemented easily on any statistical package, however by changing the database brings extra uncertainty that should be properly treated. The approach used by Wilcosky and Chambless~\cite{wilcosky1985comparison} as opposed to the log-binomial model does not suffer from convergence problems.
 
One limitation of our results is that there is no "gold standard" for choosing the best method, specially when there is a continuous explanatory variable. Since application 2 all explanatory variables are binary, the Mantel-Haenszel method could be used as the "gold standard". And we found that the prevalence ratio estimated by log-binomial, Poisson robust, marginal prevalence ratio and Schouten et al. approach showed similar estimates according Mantel-Haenszel one, showing the equivalence of the models applied. In this paper we have not explored robust methods based on quasi-likelihood estimation~\cite{lumley2006}.

In summary, we recommend to use the direct approach proposed by Wilcosky and Chambless~\cite{wilcosky1985comparison} because it is suitable for a binary response when using a variable binomial model, has no convergence difficulties and now is implemented into open source statistical software, the R package. The marginal prevalence ratio provides similar estimates as the other methods, while the conditional prevalence ratio shows the prevalence ratio for an average person in the database. If the one is interested in a particular set of controlling variables it only needed to specify the values of variables 

\bibliographystyle{ieeetr}
\bibliography{bibliog}

\newpage

\section*{Appendix}

R code to calculate marginal and conditional prevalence ratio according to  Wilcosky and Chambless~\cite{wilcosky1985comparison}. 

\begin{Schunk}
\begin{Sinput}
#####################################################
### Prevalence ratio from logistic regression models
### Author: Leo Bastos
#####################################################

PR = function( model, type="marginal", level=0.95 ){
  # model: output of a glm funtion family=binomial
  # type: "marginal", "conditional", "both"
  
  p = model$rank-1
  n = length(model$y)
  
  xx=rep(0,model$rank-1)
  
  modeloAux = update(model, x=T)
  modeloAux2 = glm(modeloAux$y ~ modeloAux$x[,-1], 
                   family=binomial(), weights=modeloAux$prior.weights)
  
  labelCI = paste( (1 + c(-level,level))/2 * 100, "%")
  auxMatrix = data.frame(PR = model$coefficients[-1], sd = NA)
  auxMatrix[, labelCI[1]] = NA
  auxMatrix[, labelCI[2]] = NA
  
  auxList = list(Conditional=auxMatrix, Marginal=auxMatrix)
  
  # Condicional
  
  xmean = apply(modeloAux$x,2, weighted.mean, w=modeloAux$prior.weights)
  
  betaAux = modeloAux2$coefficients
  varbetaAux = vcov(model)
  
  for(k in 1:p){
    ##### Conditional
    XcondAuxXk = xmean
    
    # Xk = x+1
    XcondAuxXk[k+1] = xx[k]+1
    etaxkC = drop(betaAux %*% XcondAuxXk)
    
    # P(X_k = x+1)
    PXk1C = 1/(1+exp(-etaxkC))
    # \Delta P(X_k = x+1)
    DeltaPXk1C = XcondAuxXk * PXk1C * (1 - PXk1C)
    
    # X_k = x
    XcondAuxXk[k+1] = xx[k]
    etaxkC = drop(betaAux %*% XcondAuxXk)
    
    # P(X_k = x)
    PXk0C = 1/(1+exp(-etaxkC))
    # \Delta P(X_k = x)
    DeltaPXk0C = XcondAuxXk * PXk0C * (1 - PXk0C)
    
    #### Conditional PR_Xk
    PRXkC = PXk1C / PXk0C
    
    DeltaPRxkC = (DeltaPXk1C * PXk0C - DeltaPXk0C * PXk1C) / PXk0C^2
    varPRxkC = crossprod(DeltaPRxkC, varbetaAux) %*% DeltaPRxkC
    
    auxList$Conditional[k,] = c( 
    		PRXkC,  sqrt(varPRxkC), 
    		PRXkC*exp( qnorm(c(.025,.975))*sqrt(varPRxkC)/PRXkC )
    	)
    
    ##### Marginal
    XmargAux = modeloAux$x
    
    # Xk = 1
    XmargAux[,(k+1)] = xx[k]+1   
    etaxkM = drop(XmargAux %*% betaAux)
    
    PXk1M = mean(1/(1+exp(-etaxkM)))
    
    auxVarMarg = exp(-etaxkM) / (1+exp(-etaxkM))^2
    
    DeltaXk1M = 0
    for(j in 1:n)
      DeltaXk1M = DeltaXk1M + XmargAux[j,] * auxVarMarg[j] / n    
    
    # Xk = 0
    XmargAux[,(k+1)] = xx[k]   
    etaxkM = drop(XmargAux %*% betaAux)
    
    PXk0M = mean(1/(1+exp(-etaxkM)))
    
    auxVarMarg = exp(-etaxkM) / (1+exp(-etaxkM))^2
    
    DeltaXk0M = 0
    for(j in 1:n)
      DeltaXk0M = DeltaXk0M + XmargAux[j,] * auxVarMarg[j] / n    
    
    #### Marginal PR_Xk
    PRXkM = PXk1M / PXk0M
    
    DeltaPRXkM = (DeltaXk1M * PXk0M - DeltaXk0M * PXk1M) / PXk0M^2
    varPRXkM = crossprod(DeltaPRXkM, varbetaAux) %*% DeltaPRXkM
    
    auxList$Marginal[k,] = c( 
    		PRXkM,  sqrt(varPRXkM), 
    		PRXkM*exp( qnorm(c(.025,.975))*sqrt(varPRXkM)/PRXkM )
    	)  
  }
  
  switch(type,
         marginal = return(auxList$Marginal),
         conditional = return(auxList$Conditional),
         both = return(auxList)
  )  
}

### End code ###
\end{Sinput}
\end{Schunk}

\end{document}